\documentclass[12pt,titlepage]{article}
\usepackage[pdftex]{graphicx}
\usepackage[numbers]{natbib}

\begin{document}

\title{The ``Billion Galaxy'' Cosmological HI Large Deep Survey (BiG-CHILDS)}

\author{Steven T.~Myers\thanks{National Radio Astronomy Observatory, 
    P.O.~Box O, Socorro, NM, 87801\ ({\tt smyers@nrao.edu})}, 
Filipe B.~Abdalla\thanks{Department of Physics and Astronomy, 
    University College London, Gower Street, London WC1E 6BT, UK}, 
Chris Blake\thanks{Centre for Astrophysics \& Supercomputing,
    Swinburne University of Technology, P.O.\ Box 218, Hawthorn, VIC
    3122, Australia}, \\
Leon Koopmans\thanks{Kapteyn Astronomical Institute, University of
    Groningen, P.O.~Box 800, 9700AV Groningen, the Netherlands},
Joseph Lazio\thanks{Naval Research Laboratory, 4555 Overlook Ave.\ SW, 
    Washington, DC  20375},
Steve Rawlings\thanks{Department of Physics, Oxford University, Keble Road, 
    Oxford OX1 3RH, UK}
}

\date{{\it Astro2010 Science White Paper} (2009-02-15)\\[1.5ex]
\begin{minipage}[h]{6in}
  \begin{center}
  \includegraphics[width=5.5in]{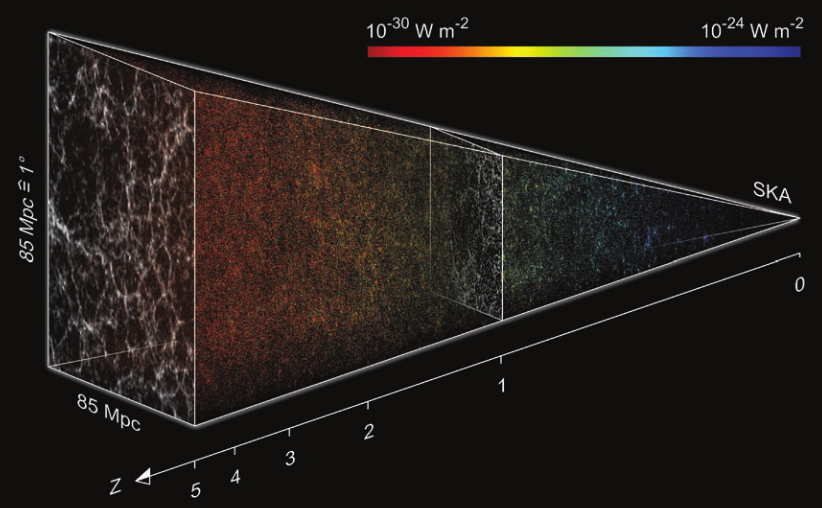}
  {\scriptsize \it Simulated SKA observing cone, courtesy D.\ Obreschkow (Oxford),
   SKADS, and the SKA project.}
  \end{center}
\end{minipage}
}

\maketitle

\centerline{\Large \sc The Billion Galaxy Cosmological HI Large Deep Survey}

\begin{quote}
{\bf Abstract:}
We outline the case for a comprehensive wide and deep survey
ultimately targeted at obtaining 21-cm HI line emission spectroscopic
observations of more than a billion galaxies to redshift $z>1.5$ over 
half the sky.  This survey provides a database of galaxy
redshifts, HI gas masses, and galaxy rotation curves that would enable a
wide range of science, including fundamental cosmology and studies
of Dark Energy.
This science requires the next generation of radio arrays, which are
being designed under the umbrella of the ``Square Kilometer Array'' 
(SKA) project. We present a science roadmap, extending to 2020 and
beyond, that would enable this ambitious survey.  We also place this
survey in the context of other multi-wavelength surveys.
\end{quote}


The first decade of the 21st century has seen cosmology build upon the
successes of the pioneering cosmic microwave background and
galaxy surveys and build a cohesive ``standard model'' for the
Universe.  In this model, Dark Energy and Dark Matter are the primary
constituents, and their densities and properties govern the evolution
of the expansion rate (Hubble constant) and the growth of structure in
the cosmic web.  As we head into the next decade, the focus is on 
probing the physics of the components of the model and testing
the limits of our understanding.  Is dark energy a quintessence fluid 
or is it a failure of general relativity?  Did inflation really set 
up the initial conditions? What is the mass of the neutrino?  
Current data is insufficient to answer fundamental questions such as these, but
next-generation cosmological surveys across the observable
electromagnetic spectrum are the necessary next steps in this enterprise as 
set forth in nationally recognized documents such as 
\cite{q2c} and \cite{detf}.  This Science White Paper focuses on the exploitation of
the 21-cm wavelength HI line emission from gas in galaxies as a probe
of the Universe from the present epoch back to a redshift $z>1.5$
through a ``billion galaxy'' Cosmological HI Large Deep Survey
(BiG-CHILDS) as a key science project for future radio astronomical facilities.


\section{Science with a Billion Galaxy HI Survey}

To enable the precise and accurate determination of the fundamental
cosmological parameters, observations must cover a large volume of the
Universe (to defeat sample variance).  A number of complementary
approaches have been proposed to meet this challenge.  For example,
the exploitation of the Baryon Acoustic Oscillation (BAO) signature 
in the galaxy angular power spectrum is a particularly promising line of
attack for dark energy studies (e.g. \cite{se03,detf}).  Other
approaches include the measurement of weak gravitational lensing
distortions in galaxy shapes, the growth rate of structure in the
Universe, and the use of standard candles (SNe) and rulers (maser
disks) in the determination of the distance scale.

\noindent\underline{\it Big Questions $+$ Compelling Opportunities:} 
A CHILD Survey would enable
astronomical observations to address key questions in cosmology
such as ---
{\bf How does rate of cosmic expansion evolve, and what physical phenomena
  control this expansion?}
{\bf What is Dark Matter?}
{\bf What is the nature of Dark Energy?}
{\bf What are the masses of the neutrinos, and how have they shaped
  the evolution of the Universe?}
A suite of cosmological surveys spanning the electromagnetic
spectrum are needed to address fundamental questions such as
these.  Furthermore, the next decade will see the fruition of key 
techniques and technologies necessary to build on the pioneering
cosmological surveys such as SDSS and provide the opportunity to
mine the cosmos and make precision measurements using truly large and 
robust samples of galaxies and AGN.


\begin{figure}[t!]
  \begin{center}
  \includegraphics[width=3.1in]{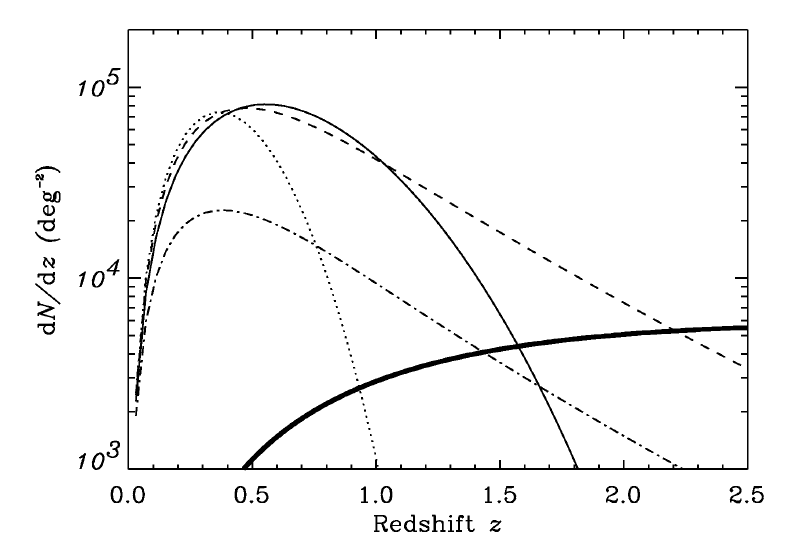}
  \includegraphics[width=2.9in]{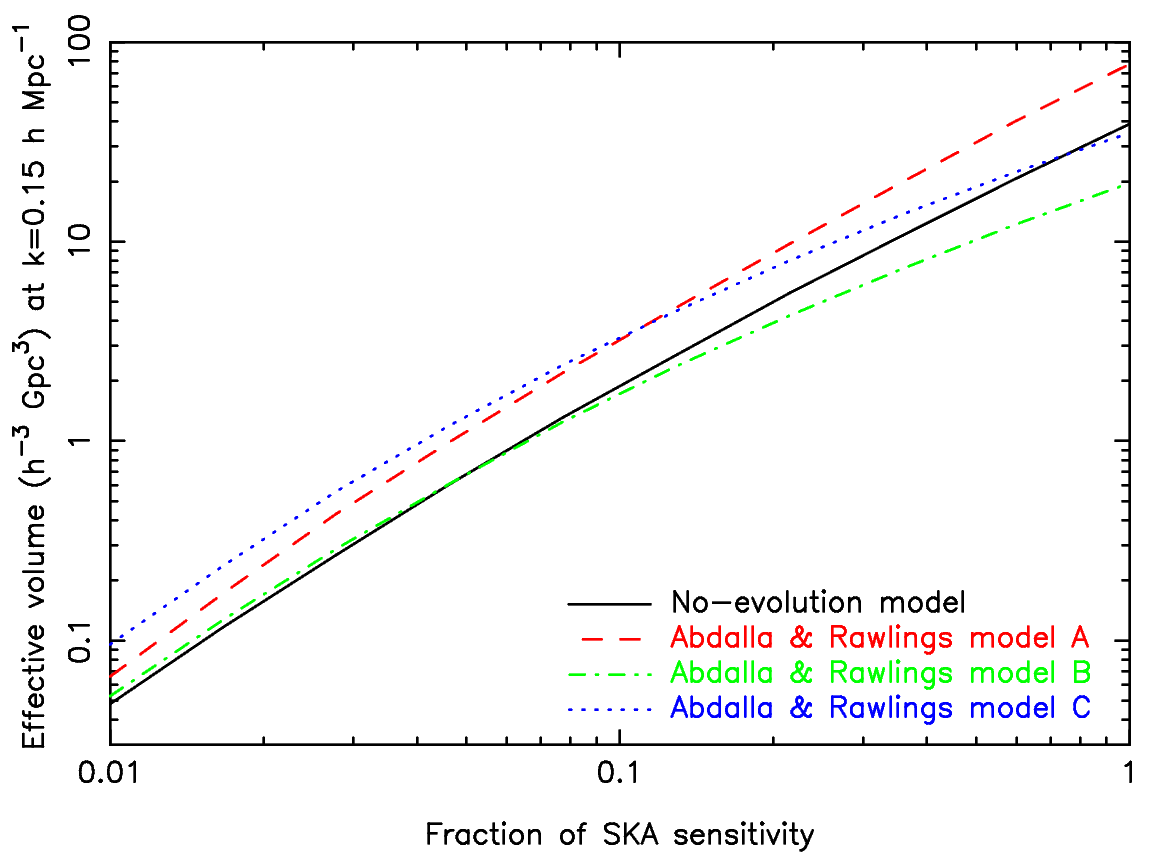}
  \caption{\label{fig:surv} 
    (a)~{\it Left:} CHILD Survey yields for various HIMF models,
    reproduced from Figure~5 from \protect\cite{AR2005}.
    The thin solid line is the favored model ``C'',
    which lies intermediate to models with no evolution (dashed)
    and evolution following dark matter haloes (dotted).
    The thick solid line shows the yield threshold needed to
    exceed cosmic variance, which for model C and a
    BiG-CHILD survey encompasses a cosmic volume out to $z\sim1.5$.
    (b)~{\it Right:} The equivalent volume on scales relevant for the BAO
    signature for a large CHILD surveys as a function of
    sensitivity (in units of a fiducial full SKA).  The
    curves are for the models in the left panel (with different
    lines, see legend).
    Note that a 10\% SKA produces a survey roughly equivalent to
    O/IR surveys projected to 2015.}
  \end{center}
\end{figure}


\noindent\underline{\it Why CHILDS?} A CHILD survey as proposed
in this white paper provides the measurement of redshifts, HI
masses, and velocity profiles for a large sample of galaxies in the
21-cm line of neutral hydrogen (HI) to $z\sim 1.5$ over a wide area of
sky.  The use of a HI survey in the context of cosmological parameter
measurement was presented in the Dark Energy Task Force (DETF)
report as a ``Stage IV'' candidate project.
The high level of performance of a BiG-CHILD BAO survey in
comparison with other DETF motivated surveys is presented in
\cite{AR2005,GlazBlake05,Tang08}. 

The future large optical/infrared (O/IR) DE surveys are 
overwhelmingly photometric in approach.
Spectroscopic surveys such as BiG-CHILDS have a key
role in the coming decade because they are particularly rich in
astrophysical and cosmological information.  For example: (1) the 
BAO signature provides robust cosmic distances and expansion
rates ($H(z)$), (2) redshift-space distortions and higher-order
clustering statistics measure the growth rate of structure, which
can be combined with the cosmic expansion history to constrain models
of dark energy and/or modified gravity, (3) higher-order clustering
and topological measures are tests of Gaussianity of the seeds of
large-scale structure and therefore the predictions of
inflation-inspired models, (4) the shape of the galaxy power spectrum
contains information on detailed physics such as the neutrino mass,
(5) spectroscopic velocities and widths can be used to determine
Tully-Fisher distances and to measure peculiar and bulk motions that
are inaccessible to purely photometric surveys,
and (6) spectroscopic redshifts are needed to calibrate photo-$z$
utilized in purely photometric surveys.

\noindent\underline{\it Survey Requirements:} 
The facility described in \cite{AR2005} would
carry out a cosmological survey competitive
with other O/IR DETF ``Stage~IV'' probes.  This
requires an array with around a square kilometer collecting area
and a field-of-view of ${\rm FoV}>10\,\rm{deg}^2$.  To optimize
survey speed\footnote{${\rm speed} \propto {\rm FoV}\times A^2/T_{sys}^2$},
one can trade off collecting area $A$ for decreased noise
system temperature $T_{sys}$ or for increased field of view.
To minimize cosmic 
variance, a comoving survey volume of $\sim 100\,{\rm Gpc}^{3}$ is needed.
Assuming a ``standard'' LCDM cosmology ($H_0=71$ km/s/Mpc,
$\Omega_M=0.27$, $\Omega_\Lambda = 0.73$) one obtains 
a comoving volume of $56\,{\rm Gpc}^{3}$ for $\sim10^4\,{\rm deg}^{2}$
with a depth of $\Delta z=0.5$ at $z=1$.  The fiducial survey would
aim to cover a hemisphere ($\sim 2\times 10^4\,{\rm deg}^{2}$).

\noindent\underline{\it Yields:} 
The counts of galaxies detected in a HI galaxy survey are set by
the HI mass function (HIMF), which is usually characterized by an
exponential times a power law \footnote{The Schechter mass function is $dN/dz = 
\phi^\ast_{HI}\,(M_{HI}/M^\ast_{HI})^\alpha\,\exp(-M_{HI}/M^\ast_{HI})$}
with a mass $M^\ast_{HI}$ characteristic of the turn-over into the exponential.
For example, \cite{AR2005} used the Parkes multi-beam HIPASS survey
HIMF of \cite{Zwaan03} to normalize the counts at $z=0$ with
$M^\ast_{HI}=3\times10^9\,M_{\odot}$ , and then
explored several evolution models.  There is a wide variation in the
counts for $z>1$ between the models due to differences between the 
$M^\ast_{HI}$ at high redshift, though $z>1$ is attainable for all
plausible models.  Using conservative model ``C'' 
\cite{AR2005} estimate a large survey could
detect $>10^5$ galaxies per square degree, giving $>10^9$ galaxies over
$10^4$ square degrees (Figure~\ref{fig:surv}a). Thus, we designate a
CHILD survey such as this the ``Billion Galaxy'' Cosmological HI Large
Deep Survey (BiG-CHILDS).

\begin{figure}[t!]
  \begin{center}
  \includegraphics[width=5.5in]{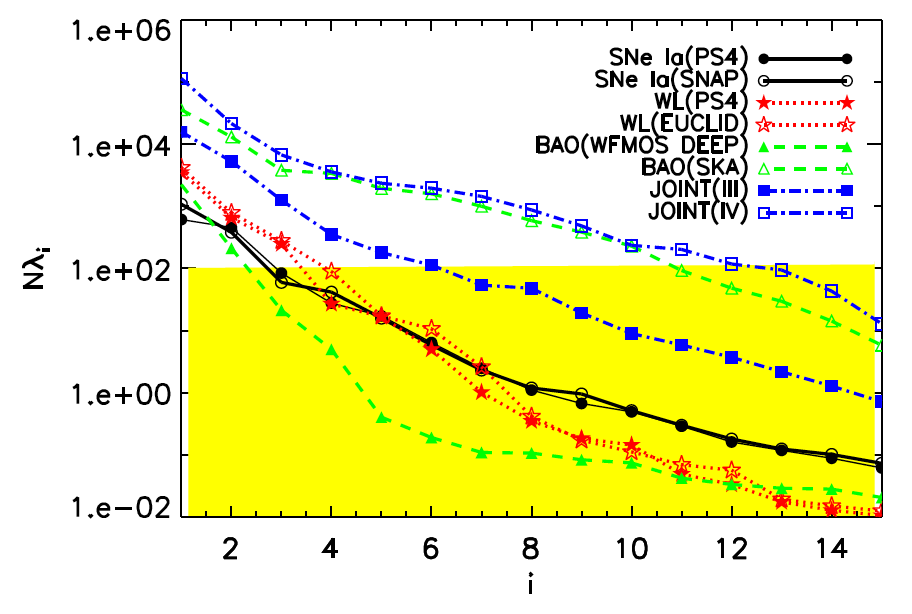}
  \caption{\label{fig:tang} 
    Dark Energy survey capabilites
    reproduced from Figure~19 from \protect\cite{Tang08}. Shown
    are the eigenvalues of the Fisher matrix as a
    proxy for signal-to-noise ratio on dark energy 
    parameterizations.  The SKA (BiG-CHILD) survey
    provides superior sensitivity by having redshifts
    for a much larger sample of galaxies than other 
    Stage III and IV experiments, allowing more complex
    Dark Energy parameterizations and equations of 
    state to be explored.}
  \end{center}
\end{figure}

Note that the billion galaxy count is not the controlling factor for
its utility for precision cosmological studies, but follows from the
need to survey a large cosmic volume to the required depth.  This 
requires the sensitivity to detect a sufficient number of galaxies at
$z>1.5$.  For a BiG-CHILD-like survey, 
\cite{Tang08} show that constraints on dark energy models
are an order of magnitude better than for other smaller volume surveys
(Figure~\ref{fig:tang}).

\section{Using the SKA for BiG-CHILDS}

The Square Kilometer Array (SKA) 
\footnote{{\tt http://www.skatelescope.org}} 
program includes a mid-frequency array component \cite{PSSKA}
that could be designed to meet these requirements.  In this context,
the cosmology science case has been previously presented by
\cite{SKAScience,RawlingsSKA,BlakeSKA} in the SKA Science Book.
The use of the SKA for the cosmological observations exemplified
by BiG-CHILDS were explored by \cite{AR2005,Tang08}.  
This requires a facility that is 40 times faster than a minimal 
``baseline'' SKA design using 3000 15-m antennas
with wideband single-pixel feeds of \cite{PSSKA}\footnote{
For reference, for a 15-m diameter antenna the ``single-pixel'' 
${\rm FoV} = 0.73\,{\rm deg}^2$ at 1.4~GHz ($z=0$),
scaling as $(1+z)^2$.}.  The necessary speedup would be attained
through a combination of increasing the survey durattion, 
operational upgrades for sensitivity and field-of-view above the
baseline design, and/or by descoping other options.   The cosmology 
science case is clearly
a driver for setting the goal for facility   
requirements, and the SKA project is focusing design and development
in this area to deal with the risk.
Note that maximizing the astronomical utility of the survey (e.g. for
dynamical studies,  
accurate position determination) requires substantial sensivity at
higher resolution ($\sim 1''$).  Thus the baseline design for SKA of \cite{PSSKA}
includes longer baselines in its proposed configuration for this reason.

CHILDS would be a nested series of 21-cm line surveys of the sky
to various depths, for which the ultimate stage would be BiG-CHILDS. 
This would be carried out commensally with other surveys for 
continuum emission from these and other galaxies, and a transient
source discovery and monitoring survey.  This program is intended
as a next generation astronomical survey with wide application beyond
cosmology, much like SDSS and its successors.  For example, another science
case for CHILDS is that of tracking 
Galaxy Evolution and structure formation, e.g. \cite{BaughSKA}.  This 
has less stringent requirements on survey speed than precision
cosmology, because smaller galaxy samples from smaller volumes can be
used.  For example, assuming the HIMF evolution model ``C'' favored by
the authors, a five year small-CHILD Survey with the baseline SKA would
yield 1 million $z>1$ galaxies over 800~${\rm deg}^2$.  Thus,
galaxy evolution studies would have extensive samples to work with,
and there would be 30--50 million $z<1$ galaxies as well.



\section{A Science Roadmap}

It is important to the health of astronomy that our current and
upcoming instruments be fully exploited to reach their science goals,
as well as to lay the groundwork for future facilities.  

\noindent\underline{\it Instrumenting the Roadmap:}
A comprehensive HI galaxy survey is a major endeavor, with capital and
operating costs for the required facility comparable to and likely
greater than that of ALMA.  A large ground-based instrument such as the SKA
will not just appear one day, fully operational and at ultimate
science capability.  Therefore, a comprehensive build-up and
build-out plan that is plausible, affordable, and above all
scientifically compelling must be developed.  This plan must involve a
large cross-section of the astronomical (not just radio) community
with a well-supported staged scientific program using the
instruments of today through to the frontiers of the next decade and
beyond.

\noindent\underline{\it Science Precursors:}
The Arecibo ALFALFA survey\footnote{{\tt http://egg.astro.cornell.edu}}
\cite{Giovanelli05}) will supplant the HIPASS
survey in anchoring the local HIMF. Dedicated use of the 
EVLA\footnote{{\tt http://www.aoc.nrao.edu/evla}}, 
a low-end enhanced ALFA, and a full 
ATA\footnote{{\tt http://ral.berkeley.edu/ata}} with
optimal performance below 1~GHz would allow us to characterize the
HIMF beyond $z=0.1$ in the coming decade.  This program, in
conjunction with the pathfinders (see below), would lead to the HI
equivalent of the SDSS.

\noindent\underline{\it Pathfinders:}
There are mid-frequency SKA pathfinders and demonstrators underway in
various locales (e.g. ASKAP, APERTIF, ATA, MeerKAT).  There is also
a US-SKA Technical Development 
Program\footnote{{\tt http://usskac.astro.cornell.edu}}
funded by the NSF.  All
of these are important to prove technology and gain experience running
SKA-like instruments.  The one that is most accessible to the US
community is the ATA.  An important goal for
pathfinder science programs is a more accurate measurement of the
$z<0.2$ HIMF, and the detection of extreme HI galaxies and also
constraint on the HIMF (e.g. through stacking) out to $z\sim0.5$.

\noindent\underline{\it Risks and Mitigation:} 
The design requirements placed on a SKA in order to
attain the sensitivities needed to carry out a BiG-CHILD Survey
are extreme compared to other SKA science drivers, and is thus
expensive.  The detection of cosmological HI is what originally
suggested that a square kilometer of collecting area was needed.  
The SKA project is exploring the options to achieve these goals,
including an upgrade path enhancing the field-of-view and
sensitivity, or descoping by trading off other aspects (such as upper
frequency limit).  A more modest small-CHILDS, more on the scale
of near-future ``Stage III'' DETF experiments, would start
the SKA on the road to the bigger survey while providing first 
science.  A 10\% Phase I SKA, for example, would have the capability
to carry out cosmological surveys equivalent to the O/IR surveys projected
for 2015 (see Figure~\ref{fig:surv}b).

\noindent\underline{\it Survey Science:} 
The concept of a Radio Synoptic Survey Telescope (RSST)
\footnote{{\tt http://www.aoc.nrao.edu/$\sim$smyers/rsst}}
has been developed to encapsulate the mid-freqency SKA science case.
The proposed CHILD Survey is but one key
component of the science opportunities afforded by a SKA as RSST.  
Even CHILDS itself would provide additional
information to the spectroscopic redshift survey, such as very 
deep continuum polarimetric imaging of galaxies and AGN.  For example,
galaxy imaging could be used as a weak-lensing probe of cosmology and
dark energy complementary to the BAO signature (e.g. \cite{BlakeSKA}).
Also, detection of HI absorption signatures against higher redshift 
AGN \cite{KanekarSKA} also provides an important probe of galaxies, 
much like damped Lyman Alpha absorbers in the O/IR bands.
The staged approach would allow the various CHILDren to build upon
each other in all these science areas.

\noindent\underline{\it Complementarity:} 
The BiG-CHILD Survey is essentially a galaxy
redshift survey, and thus will provide a hydrogen-filtered view of
the cosmic web.  Note that future photometric surveys at other wavelengths, 
such as those proposed in the O/IR bands for PanSTARRS and
LSST, will also yield around a billion galaxies, and
cross-correlation with BiG-CHILDS would be of great utility for both
cosmology and galaxy studies.  For example, BiG-CHILDS galaxies can be
used to calibrate the photometric redshifts used in these
non-spectroscopic surveys.

\noindent\underline{\it Alternatives:}
The science goal of measuring the BAO signature in the HI galaxy power
spectrum could also be approached through a very low resolution high
sensitivity interferometer array \cite{Wyithe08}.  This would be much
like scaling up proposed HI EOR experiments, at potentially
lower cost than a CHILDS-capable SKA.  However, unlike BiG-CHILDS,
this experiment does not deliver a combined continuum and
spectroscopic sample of galaxies that is useful for more general 
astronomical science.

\noindent\underline{\it Timeline:} 
The ultimate facility required to carry out BiG-CHILDS would only be
possible in the post-2020 timeframe.  However, significant staged
CHILDren surveys will be carried out in 2010--2015 through science
precursor programs on facilities such as Arecibo, ATA, EVLA, and GBT.
In the latter half of the decade, the international SKA pathfinders
such as an expanded ATA and ASKAP will take this science to the next
stage.  Finally,
construction on a first phase of the SKA later in the decade could
start to yield HI galaxy samples at $z>0.5$ around 2020.

\noindent\underline{\it Uniqueness:} The BiG-CHILD survey would be a unique resource for
astronomy, astrophysics, and cosmology.  It is a simulataneous 
spectroscopic and photometric (continuum) radio survey.  With a 
yield of more than a billion redshifts, it would be the ultimate
ground-based cosmological redshift sky survey, and an important
counterpart to space-based spectroscopic missions targeting
similar-sized samples.  The unprecedented velocity information
will enable key tests of GR, redshift space distortion probes of
cosmology, large-scale flows, and the nature of initial velocity
perturbations. Radio surveys and techniques do have some
advantages over O/IR surveys, such inherently
wide instantaneous field-of-view ($>1\,{\rm deg}^2$), and 
insenstitivity to galactic obscuration.  However, the greatest 
science utility comes from the combination of all multi-wavelength
surveys, including BiG-CHILDS.

\noindent\underline{\it Related Development:} In addition to the substantial technology
development needed to construct the SKA, significant advances in 
theory, modeling, and data analysis algorithms and techniques is also
required.  Extracting a billion-galaxy catalog from the huge raw data
volumes that BiG-CHILDS will produce is a daunting challenge.
Furthermore, carrying out the cosmological and astrophysical
simulations needed as input to model the theoretical part of the 
parameter extraction will also require development during the decade.
The staged science approach will allow this to grow with the CHILDren
surveys.  Support for next-generation algorithmic and computational
infrastructure will be critical to enable the science from surveys
such as BiG-CHILDS.

\section{Conclusions}

The Billion Galaxy Cosmological HI Large Deep
Survey, in conjunction with commensal continuum and transient-response
surveys on the same instrument, would comprise a truly
transformational science program.  The BiG-CHILDS key project would
provide superb constraints on cosmological parameters such
as the dark energy equation of state.  A staged survey suite leading
from the facilities of today to the SKA beyond 2020 would enable this
exciting opportunity.




\bibliographystyle{unsrt}

\end{document}